\magnification=1200
\baselineskip=20 pt

\def\sigmamunu{\sigma^{\mu\nu}}
\def\fmunu{F_{\mu\nu}}
\def\zmunu{Z_{\mu\nu}}

\centerline{\bf Constraining new physics in the Tev range by
the recent BNL measurement of $(g-2)_{\mu}$}

\vskip .5 true in

\centerline{\bf Uma Mahanta}

\centerline{\bf Mehta Research Institute}

\centerline{\bf Chhatnag Road, Jhusi}

\centerline{\bf Allahabad-211019, India}

\vskip 1 true in

\centerline{\bf Abstract}

In this paper we study the implications of the recent high precision 
measurement of $(g-2)_{\mu}$ by BNL [1] on new heavy physics beyond the SM
in a model independent way.
We find that if the new physics responsible for the muon anomaly is due to
 d=6 direct operators then they could arise from the following
follwing three broad classes of new physics
 a) new particles in the few hundred Gev range  with weak gauge coupling b)
 strongly interacting particles  and resonances in the few Tev range and
c) massive Kaluza-Klein modes of the graviton in the Tev range 
with couplings to SM particles of the order of ${E\over (Tev)}$.

\vfill\eject

The effect of new heavy physics beyond the SM  appearing above some high
energy scale $\Lambda$  at energies much small compared to $\Lambda$
can be expressed by non-renormalizable operators constructed out of SM fields.
These operators can be expressed in a systematic power series expansion in 
${1\over \Lambda}$. The structure of these operators is completely
determined by a) the fields that are dynamical at the relevant energy scale
b) the residual gauge symmetry at scales much small compared to $\Lambda$
and c) the global symmetries respected by the low energy theory.

The muon $(g-2)_{\mu}$ collaboration has reported a new improved 
measurement of the positive muon anomaly [1]

$$a_{\mu}(expt)=11659202(14)(6)\times 10^{-10}.\eqno(1)$$

The value currently expected in the SM is [2]

$$a_{\mu}(SM)=11659159.6(6.7)\times 10^{-10}.\eqno(2)$$

The world-average experimenta value of $a_{\mu}$ shows a discrepancy of
2.6 $\sigma$ from the SM value [1]

$$\delta a_{\mu}=a_{\mu}(expt)-a_{\mu}(SM)=43(16)\times 10^{-10}.\eqno(3)$$

The new measurement of the muon anomaly by BNL  has produced quite some
interest and activity in this area. The discrepancy between the SM and 
the experimental value reported by BNL has been used to put bounds on the
unknown parameters in a variety of new physics scenario namely extra 
gauge bosons, exotic fermions, compositeness, supersymmetry and leptoquarks
[3].

In this report we shall study the effect of new heavy phyics appearing
at some high energy scale $\Lambda$ on the anomalous magnetic moment of 
the muon. Such effects can be expressed by non-renormalizable operators
constructed out of SM fields. The operators must be invariant under
the SM gauge group $SU(3)_c\times SU(2)_l\times U(1)_y$ which is the
relevant
gauge symmetry below $\Lambda$. We shall assume that the SM gauge symmetry
is linearly realized on the SM fileds. This will correspond to an elementary
 or a light composite higgs scalar.
It then follows that
the lowest dimension operator
invariant under the SM gauge group that
contributes to the muon magnetic moment anomaly is six. Here we shall
consider two such operators
that contribute directly to $a_{\mu}^{np}$ 
 and determine the lower bound on the scale
associated with them from the new physics contribution to muon anomaly
reported by BNL. A detailed discussion of effective Lagrangian analysis 
of muon anomlay can be found in Ref[4].
 The two direct operators of dimension six that contributes
to $a_{\mu}$ are [5]

$$\eqalignno{O_1&=({\bar l}\sigmamunu\tau_{a}\mu_R)\phi W^a_{\mu\nu}
+h.c.\cr
&=-{1\over {\sqrt 2}}({\bar \mu}\sigmamunu \mu ) (c_w \zmunu +s_w
\fmunu ) (v+H)+..&(4)\cr}$$

and

$$\eqalignno{O_2&=({\bar l}\sigmamunu \mu_R)\phi B_{\mu\nu}\cr
&={1\over {\sqrt 2}}(c_w\fmunu -s_w\zmunu )(v+H)+..&(5)\cr}$$

The low energy effective Lagrangian relevant for us  is therefore

$$L_{eff}={C_1\over \Lambda^2}O_1+{C_2\over \Lambda^2}O_2.\eqno(6)$$

The coefficients $C_1$ and $C_2$ can arise from three broad classes
of new physics a) weakly coupled gauge theory b) strongly coupled
gauge theory c) theories in extra space-time dimensions.
 The size of the 
coefficients will depend  upon from which kind of new physics it arises.
a.) If the new physics that give rise to $O_1$ and $O_2$ is 
a weakly coupled gauge theory
 then the coefficients $C_1$ and $C_2$ can be estimated 
by explicitly
 evaluating the loop diagrams made of virtual states of new heavy
particles perturbatively. Typically ${s_w\over \sqrt {2}} C_1$
and ${c_w\over \sqrt {2}} C_2$ 
are expected to be of the order of $e{g^2\over 16\pi^2}\xi$ where
g is some weak coupling that appears at the scale $\Lambda$. $\xi\approx
{m_F\over v}$ where $m_F$ is the mass of some internal fermion line. The
parameter $\xi$ carries the information that the operators $O_1$
and $O_2$ break chiral symmetry. The muon anomaly due to $O_1$ in the
weakly coupled scenario is therefore given by

$$a_{\mu}^{np}= {m_{\mu}m_F\over\Lambda^2}
{g^2\over 8\pi^2}.\eqno(7)$$

We would like to note that the interaction that gives rise to 
anomalous magnetic moment of muon will also contribute to the muon mass.
The shift in the muon mass will be  given by 
$\delta m_{\mu}\approx {g^2\over 16\pi^2} m_F \ln{M\over m_{\mu}}$
where M is the mass scale which gives the dominant contribution 
to the loop integral. 
In the weakly coupled case the coupling g must be small enough so that
$\delta m_{\mu}<< m_{\mu}$ and the muon mass is protected
from receiving large radiative corrections from the new physics scale.
 The best known example of weakly
coupled new physics that also satisfies the criteria of naturalness
 is the the supersymmetric version of the SM. In such a scenario
 the muon gets its mass from yukawa coupling to one of the higgs doublets.
However the dynamics that gives rise to the phenomenological
 yukawa couplings is assumed 
to take place at an absurdly high energy, certainly much higher than
the mass scale of the weakly interacting new particles that give
rise to the  muon  anomaly.
 Elementary leptoquarks constitute
a non-supersymmetric example weakly coupled new physics.
 If the muon anomaly
is due to a second generation leptoquark then we have $m_F=m_c$
(mass of charm quark)
and $\delta m_{\mu}\approx {g^2\over 16\pi^2} m_c \ln {M\over m_c}\approx
5.1 Mev<< m_{\mu}$ if the coupling g of the leptoquark
to quark-lepton pair is of the
order of electromagnetic coupling. Further in this case the muon anomaly
is given by $a_{\mu}^{np}\approx {g^2\over 8\pi^2} {m_{\mu} m_c\over
\Lambda^2}$. The new BNL value of the muon anomaly is important because
of two reasons. Firstly the average value of the anomaly is large
(2.6 $\sigma$ effect). Secondly the error in the new value is one 
third of the combined previous data. Both these factors can be
taken into account by determining 95\% CL limits on $\Lambda$.
For the leptoquark case we find that for $g\approx e$, $\Lambda$\
must satisfy the following bounds: 160 Gev $\le \Lambda \le $378 Gev.
In contrast the previous data [2] ($\delta a_{\mu}\approx 45(46)\times
10^{-10}$) would have given us a central value of 350 Gev and
a lower limit of 105 Gev for $\Lambda$. Clearly the new BNL value
allows a much more precise determination of the scale of new physics.
In general for weakly coupled scanario (due to the small
coupling and loop suppression  factor ) we expect 
${s_w\over \sqrt {2}} C_1$ to be much smaller than ${m_{\mu}\over v}$.
Assuming a typical suppression factor of .01-.04 we expect new particles 
to appear in the few hundred Gev range with couplings of the order
of electromagnetic coupling or even weaker.
 
b.)In the strongly coupled case ( as for example in composite models )
on the other hand,
 the underlying physics that give rise
to the mass of the muon appears at the scale relevant for the muon anomaly
 itself. This happens for example in extended technicolor models.
 Hence the results
for this case can be obtained by setting the expression for $\delta m_{\mu}$
or $vC_1$ given above
equal to $m_{\mu}$. In that limit the expression for the muon anomaly
due to $O_1$ becomes $a_{\mu}^{np}\approx                         
{m_{\mu}^2\over \Lambda^2}$. To justify that this expression is correct
consider a 
nonabelian gauge theory where the small muon mass arises from the strong 
binding of very massive preons then the anomalous magnetic moment
of the muon is expected to be of the order of $\delta \mu\approx e {m_{\mu}
\over \Lambda^2}$ [6].
 The mass of the muon must appear in the numerator since 
a non zero $\delta \mu$ ( anomalous magnetic moment) implies
 chiral symmetry breaking in the light
composite muon. The above expression for $\delta a_{\mu}$
also arises in theories where the muon gets its mass from extended
technicolor interactions (ETC) [7]. To see that consider the loop diagram
with an exchange of ETC gauge boson that gives rise to an anomalous
 magnetic moment of the muon. A simple calculation shows 
that the anomalous magnetic moment would be given by

$\delta \mu\approx e {g^2_{etc}\over 16\pi^2}{\langle \bar {T}T\rangle \over 
M^4_{etc}}$. Here $g_{etc}$ is the  ETC gauge coupling, $M_{etc}$
is the mass of the extended technicolor gauge boson and
$\langle \bar {T}T\rangle$ is the technifermion condensate renormalized at
$M_{etc}$. The exchange of the same
 ETC gauge boson will also generate the muon 
mass and will be given by $m_{\mu}\approx  
{g^2_{etc}\over 16\pi^2}{\langle {\bar T}T\rangle \over 
M^2_{etc}}$. From these two equations it follows that the 
muon anomaly due to ETC interaction will be given by
$a_{\mu}^{np}\approx 
{m_{\mu}^2\over M^2_{etc}}$.
Using the new BNL value for muon anomaly we get the following 95\%
CL range for $\Lambda$ in the strongly coupled scenario:
1.2 Tev$\le \Lambda \le $ 3.2 Tev. The ultimate goal of the experiment
is to reduce the error to $\pm 4\times 10^{-10}$ about a factor
of 3.5 times better than the present result. Even the inclusion of 
already existing data from 2000 run would ipmrove the statistical error by 
a factor of 2. If the central value and other errors are unaffected, 
the 95\% CL bounds will become: 1.3 Tev $\le \Lambda \le $2.3 Tev.

c.) Theories in extra space-time
dimensions: Recently theories in extra dimension have 
been proposed to explain the hierarchy problem [8].
 In this section we shall
estimate the coefficient $C_1$ associated with the direct operator $O_1$
assuming that it arises from a higher dimensional model. In these models
the SM fields are assumed to be localized on a 3 brane but gravity is
allowed to propagate in the bulk. From the point of view of an
observer in the visible four dimensional world the effect of having gravity
in the bulk  is described by a tower of Kaluza-Klein (KK)
 modes of the graviton with a level spacing of a few Tev.
The zero mode is the usual graviton and it couples to the SM fields
with a strength proportional to ${E\over M_p}$ where $M_p$ is the Planck
mass. But the higher KK modes lie in the Tev range and they couple to SM 
fields with a strength proportional to ${E\over \Lambda}$ where
$\Lambda$ is of the order of a Tev. The couplings of the KK modes of 
the graviton to SM particles are therefore strong for energies in the
Tev range.
 Consider now a muon self
energy diagram with an exchange of KK graviton. Attach  a photon 
and a higgs field to the muon line. On integrating over  the fluctuations
of all the massive
KK graviton modes such a diagram will generate the operator $O_1$.
Pulling out the photon momentum out of the integral and evaluating
the loop integral (which receives largest contribution from
 loop momenta of the order
of the cut off $\Lambda$)
 we find that

$$C_1 O_1={e\over 16\pi^2}{m_{\mu}\over v}
\ln {\Lambda^2\over
m_{\mu}^2} \bar {\mu} \sigma_{\mu\nu}\mu F^{\mu\nu} h+...\eqno(8)$$

Hence ${\sqrt {s_w}\over 2} C_1\approx {e\over 16\pi^2}{m_{\mu}\over v}
\ln {\Lambda^2\over m_{\mu}^2}$. This will generate a muon anomaly
of the order of $a_{\mu}^{np}\approx .24 {m_{\mu}\over\Lambda^2}\approx
2.4\times 10^{-9}$. Here we have assumed that $\Lambda\approx $ 1 Tev.
This is of the right order of magnitude to give rise to the observed
BNL muon anomaly. Hence extra dimension scenarios with massive KK
modes of the graviton that couple to SM fields with a strength
inversely proportional to the Tev scale, can also give rise to the
direct operators with coefficients of the right size to generate the BNL
muon anomaly

The operators $O_1$ and $O_2$ are the only two operators among the
d=6 operators which contribute directly to $a_{\mu}^{np}$ at the tree level.
However there are many d=6 operators that contribute indirectly to
$\delta a_{\mu}$ through loops. 
Here we shall consider only those indirect operators (with an
elementary or  a light composite higgs scalar)
 whose effects on the
muon anomaly have not been considered in Ref.[9].

a) Effect of indirect operators made of gauge bosons and scalars:
Consider the two operators [5]
 $O_3=(\phi^+D_{\mu}\phi ) (D_{\mu}\phi^+\phi )$
and $O_4=(\phi^+\phi )(D_{\mu}\phi^+ D_{\mu}\phi)$.
These operators cause $O({v^2\over \Lambda^2})$ mixing between
$W_{3\mu}$ and $B_{\mu}$. They also shift
 the physical W(Z) boson masses by an amount $\delta m^2_
{w(z)}\approx {v^2\over \Lambda^2} m^2_{w(z)}$. If this operator
appears in the low energy effective Lagrangian with a coefficient
of the order of one, then it would cause a shift in the $\rho$
parameter by an amount $\delta \rho_{new}\approx -O({v^2\over
\Lambda^2})$. The LEP constraint $| \delta \rho_{new}|\le
$.4 \% implies that ${(v^2\over \Lambda^2})$ can be at most of the
order of one percent. If the operator $O_3$ is introduced on
a weak gauge boson line of a loop diagram that contributes to
$a_{\mu}^{ew}$ then the change $a_{\mu}^{np}$ due to $O_3$
will be given by $a_{\mu}^{np} \approx O({\delta m^2_z\over m^2_z})
a_{\mu}^{ew}\approx {v^2\over \Lambda^2} a_{\mu}^{ew}\approx 10^{-11}$.
This contribution is much smaller than the ultimate precision 
$ 4\times 10^{-10}$ that the
BNL collaboration aims to achieve. The $O({v^2\over \Lambda^2})$ mixing
between   $W_{3\mu}$ and $B_{\mu}$ will also affect the direct operators
$O_1$ and $O_2$. When $W_{3\mu}$ and $B_{\mu}$ are
 expressed in terms of physical states $Z_{\mu}$ and $A_{\mu}$ the
shift in $a_{\mu}^{np}$ will be of order 
${v^2\over \Lambda^2}a_{\mu}^{direct}$. Here 
$a_{\mu}^{direct}$ is the contribution to $a_{\mu}^{np}$ due to
$O_1$ and $O_2$. Hence the effect of this operator on the muon anomaly is
 too small to be observed with the present precision.

b)Effect of indirect operators made of gauge bosons, fermions and scalars:
Consider the operator [5]
$$\eqalignno{O_5 &=(\bar {l}_{\mu}D_{\mu} \mu )D^{\mu}\phi \cr
&=-{\imath v\over 2\sqrt 2} (g^2+g^{\prime 2})^{1\over 2} Z_{\mu}
\bar {\mu}_l\partial ^{\mu} \mu_r +...&(9)\cr}$$
Since the operator $O_4$ breaks the chiral symmetry of the muon the
dimensionless coefficient associated with it in the low energy effective
Lagrangian
 must be proportional
to the Yukawa coupling ${m_{\mu} \over v}$ of the muon. This would guarantee
that in the limit of vanishing muon mass the operator $O_4$ vanishes
and the chiral symmetry of the muon is recovered.
Hence the relevant term in the effective Lagrangian becomes
$${C_4\over \Lambda^2}O_4\approx -{\imath v\over 2\sqrt 2}
(g^2+g^{\prime 2})^{1\over 2}Z_{\mu}\bar \mu_l\partial ^{\mu}
\mu_r+...\eqno(9)$$. 

Consider now a loop diagram
with Z boson exchange that contributes to $a_{\mu}^{ew}$ in the SM.
Replace one of the SM vertices of Z by the above nonrenormalizable
effective vertex. The resulting diagram will give the contribution
 of $O_4$ to $a_{\mu}^{np}$. Pulling out the photon
momentum out of the integral
 and evaluating the resulting integral we get
$$a_{\mu}^{np}\approx {m_{\mu}^2\over \sqrt {2}\Lambda^2}{e^2\over
16\pi^2 c_w^2}\ln{\Lambda\over m_z}.\eqno(10)$$.
This contribution is of the order
of $.001 {m_{\mu}^2\over \Lambda^2}$ and hence much smaller than the
direct contribution considered in this paper.

The dimension less coefficient associated with the chirality flipping
direct operators $O_1$ and $O_2$ must be order of the yukawa coupling of
the muon and hence these operators will not produce any significant
effect at a high energy $\mu^+\mu^-$ collider. However the new physics
that give rise to the chiral symmetry breaking operators can also give rise
to chiral symmetry conserving operators. The coefficient of these operators
in the low energy effective Lagrangian could be of order one.
Consider for example the operator $O_6=i[ (\phi^+D_{\mu} \phi)-
(D_{\mu}\phi^+ \phi)]\bar{l}
\gamma^{\mu}l$. Besides shifting the Z coupling to LH muons this
 operator could also give  rise to anomalous contributions
to the process $\mu^+\mu^-\rightarrow hZ$. Although the contribution of this
indirect operator to the muon anomaly is suppressed [9] compared to that of the
direct operators, its collider signatures are stronger than that of
the direct operators. It will be interesting to study the collider
implications of this and other d=6 indirect operators which are not chirality
suppressed for the values of $\Lambda$ presented in this paper.
 This would help in revealing the complementary nature
of the new physics responsible for the muon anomaly.
In fact this idea was used by Eitchen et al in Ref[10] to propose
the study of four fermion contact interactions at a high energy
$e^+e^-$ collider to determine the bounds on the compositeness scale
$\Lambda$. Until then  the muon anomaly was considered as providing
the best bound on the compositeness scale.
Motivated by this complementary search strategy we have done some
rough estimates of the effect of the operator $O_5$ on the process
$\mu^+\mu^-\rightarrow hz$. We find that at $\sqrt{s}$=500 Gev,
$m_h=150 Gev$ and $\Lambda $=2 Tev, $\sigma_{total}$=80 fb
if $O_6$ interferes constructively with the SM contribution.
This is to be compared with the SM contribution of 52 Gev.
Hence unless $\Lambda$ is much higher we could expect to see large
 new physics effects in the process $\mu^+\mu^-\rightarrow hZ$.
Similarly the operator $O_5$ can make important new physics contributions
to higgs production on resonance at a 500 Gev $\mu^+\mu^-$
 collider.

In conclusion in this paper we have 
shown that the scale associated with the direct operators that can
 explain the BNL muon anomaly can naturally arise from the following
three distinct scenarios:  
 a) weakly coupled
gauge theories b) strongly coupled gauge theories and
c) theories in extra space time dimension. 
Remarkably all the three scenarios namely weak scale
supersymmetry, technicolor and theories in extra dimensions
provide solutions to the hierarchy problem.
 If the direct operators  arise from
a  weakly coupled  underlying gauge theory, the scale of new 
physics typically  turns out to be a few hundred Gev. Here we expect
weakly coupled new particles (leptoquarks or supersymmetric partners
of the SM particles)
 with a mass of the order of few hundred Gev.
On the ohther hand if $O_1$ and $O_2$ arise from a 
 strongly coupled gauge theory, the scale of new physics turns out to be
a few Tev. In this scenario we expect strongly coupled new particles
( technihadrons and technimesons )
and   other resonances with a mass in the Tev range.
Finally if the the direct operators arise from extra dimension theories
with gravity living in the bulk then we expect to see the massive KK modes
of the graviton in the Tev range which couple to energy momentum tensor
of SM fields with only
Tev scale suppression. Interestingly all the three scenarios for new
physics will be accessible at LHC and other future colliders for
detailed study.

\centerline{\bf Acknowlegdement}

The auhor would like to thank Dr. M.  Einhorn and Dr. K. Lane for
comments which led to the revised version of the manuscript.

\centerline{\bf References}

\item{1.}H. N.  Brown et al, Muon g-2 collaboration, hep-ex/0102017.

\item{2.} A. Czarnecki and W. Marciano, Nucl. Phys. (Proc. Supp) B 76, 245
(1999).

\item{3.} A. Czarnecki and W. Marciano, hep-ph/0102122; K. Lane, 
hep-ph/0102131; L. Everett, G. L. Kane, S. Ringolin, and L. T. 
Wang, hep-ph-0102145; J. Feng and K. Matchev, hep-ph/0102146; 
E. Baltz and P. Gondlo, hep-ph/0102147; U. Chattopadhyay and P. Nath, 
hep-ph/0102157; U. Mahanta, hep-ph/0102176; D. Chowdhury, B.
Mukhopadhyaya and S. Rakshit, hep-ph/0102199.

\item{4.} C. Arzt, M. Einhorn and J. Wudka, Phys. Rev. D 49, 1370 (1994).

\item{5.} W. Buchmuller and D. Wyler, Nucl. Phys. B 268, 621 (1986).

\item{6.} R. Barbieri, L. Maiani and R. Petronzio, Phys. Lett. B
96, 63 (1980); S. J. Brodsky and and S. D. Drell, Phys. Rev. D 22,
2236 (1980).

\item{7.} E. Eitchen, K. Lane and J. Preskill, Phys. Rev. Lett. 45,
225 (1980).

\item{8.} N. Arkani-Hamed, S. Dimopoulos and G. Dvali, Phys. Lett.
B, 429, 263 (1998); I. Antoniadis, N. Arkani Hamed, S. Dimopoulos
and G. Dvali, Phys. Lett. B, 436, 27 (1998); L. Randall and R. Sundrum,
Phys. Rev. Lett. 83, 3370 (1999); L. Randall and R. Sundrum, Phys. 
Rev. Lett. 83, 4690 (1999).

\item{9.} M. B. Einhorn and J. Wudka, hep-ph/0103034.

\item{10.} E. Eitchen, K. Lane and M. Peskin, Phys. Rev. Lett. 50,
811 (1983).

\end